# Symmetrical field effect and slow electron relaxation in granular aluminium

T. Grenet

*LEPES-CNRS, BP 166, 38042 Grenoble cedex 9, France*
*E-mail: thierry.grenet@grenoble.cnrs.fr*

**Abstract.** Conductivity and field effect measurements in thin insulating Al granular films are reported. The occurrence of a symmetrical field effect and of very slow conductance relaxations is demonstrated. They are identical to the electron glassy behaviours already reported in insulating indium oxide thin films. The results suggest that the phenomena are quite general. The study of structurally discontinuous samples should help to understand the origin and mechanism of the glassy behaviour.

In a recent series of papers [1-6] remarkable glassy phenomena have been demonstrated in the conductance of insulating disordered indium oxide (InOx) films. Field effect measurements performed at low temperature using MOS-like devices exhibited an unexpected symmetrical effect, the InOx channel conductance G increasing when small gate fields of either sign were applied. Moreover these G(Vg) characteristics (where Vg is the gate voltage) were shown to relax very slowly (over hours) after Vg or T were changed. These effects have been attributed to a glassy state of the electrons, promoted by the disorder and the electron correlations. A symmetrical field effect had also been reported previously in Au granular films [7], and more recently in ultrathin metal-Ge bilayers [8, 9]. In these works slow relaxation was also briefly mentioned, however these exciting phenomena have only been studied in detail in InOx thin films so far. If mainly due to disorder and electron-electron interactions, they should be present in any other disordered insulator. Granular metals are particularly interesting candidates to study them. They can be on either side of the metal to insulator transition. One can expect slow dynamics because of their high electron densities, as it was shown in InOx films that the higher the carrier density is, the slower the relaxations are. Moreover, one can in principle control the microstructure of the granular samples (grain size, intergrain coupling), which offers an efficient tool to track the microscopic ingredients and mechanism of the phenomena. In this paper I study the field effect in granular aluminium films, and show that the very same glassy effects as those observed in InOx also exist in that system, which opens a promising route for their investigation and understanding.

The system under study is granular Al obtained by evaporating Al in a reduced oxygen pressure. The samples are MOS-like devices consisting of an Al gate, an $Al_2O_3$ gate insulator and a granular Al channel. They were prepared by e-beam evaporation in a high vacuum chamber (base pressure ~$10^{-7}$ mbar) using the following procedure. An Al gate was first deposited on polished sapphire substrates, followed by a 1000Å thick insulating $Al_2O_3$ film. The 200Å thick channel to be studied was then obtained by evaporating Al in a controlled oxygen pressure. Typical conditions were an evaporation rate of 2 Å/s and P($O_2$) of a few $10^{-5}$ mbar. Finally Al contacts were deposited on the sample in vacuum, with a geometry such that R~R/80. The whole process including masking was performed without opening the chamber. The microstructure of granular Al films prepared by evaporating Al in $O_2$ atmospheres is well known [10]. They consist of Al grains separated by oxide layers. When the oxygen pressure is increased, one gets more isolated and smaller grains, with a broader size distribution. By controlling the (oxygen pressure / Al deposition rate) ratio, one can control the sample's resistance and location with respect to the metal to insulator transition. Once removed from the preparation chamber, the samples were immediately mounted on the sample holder and put in the cryostat. A series of samples was prepared with different (oxygen pressure/Al evaporation rate) ratios, ranging from metallic to highly insulating. No field effect could be detected in the (marginally) metallic films (like a sample with R (4K)/R (300K)~2 and R (4K)~7 kΩ). However a clear field effect and slow relaxation phenomena were observed in the two insulating samples which were measurable at 4 K. In the following I only present the results obtained on the less insulating one (R (4K)~280 MΩ), but similar effects were also observed in the second one.

The channel conductance measurements were performed in a He4 cryostat, imposing a voltage across the sample and measuring the current using an EGG5209 lock-in amplifier associated to a home made current amplifier. Owing to the high sample's impedance, the measurements were performed at low frequency (generally 4.5 Hz) and it was checked that the results do not change significantly when the frequency is changed. Potential artefacts due to leakage currents through the gate insulator when Vg is applied should be negligible. The gate insulator was tested at 4K by applying Vg values in the range used for the field effect experiments. No leakage current could be detected within the DC measurement sensitivity which was 1 pA (which corresponds to 10% of the amplitude of the field effect we are interested in). Moreover, since the sample's G measurements are AC, the lock-in amplifier filters eliminate any contribution from these possible sub-picoamp leakage currents.

Fig.1 shows the T dependence of the resistance of the R (4K)~280 MΩ sample, which increases by four orders of magnitude from room temperature to 4K. No simple fit (e.g. power law, exponential, variable range hopping (VRH) type) holds on the whole temperature range. Below 30 K, the data precisely follow an Arrhenius activated law as is shown in the inset of Fig. 1. This is at variance with the general observation of the Efros-Shklovskii VRH type dependence in insulating granular metals, and the reason for this discrepancy is not clear yet.

I concentrate on the field effect measurements. In Fig. 2 I show a typical G(Vg) curve for the sample equilibrated for more than 20 hours at T=4K and Vg=0. One sees a clear symmetrical "dip", with no significant anti-symmetrical component. Its amplitude amounts to δG/G~2%, and its FWHM corresponds to a charging of the gate-sample capacitor of $\Delta Q \sim 5 \cdot 10^{11}$ e/cm$^2$ ($\Delta Q = C \Delta V_{g_{1/2}}$ with $C=\varepsilon_\circ \varepsilon_r(Al_2O_3)S/t$, S=1cm$^2$, t=10$^3$ Å) . Changing the Vg value induces a slow relaxation of the G(Vg) curve, the initial dip being progressively filled while another one appears centred on the new gate voltage value. This is shown in Fig.3, where it can be seen that this evolution is extremely slow, as a trace of the initial dip remains more than 10 hours after Vg was changed! Actually the rate at which the first (Vg=0) dip disappears depends on the duration of the equilibration of the sample. In Fig. 4 are shown ageing experiments similar to the ones performed by Vaknin et al. [5] on InOx. In these experiments, once equilibrated at Vg=0, the sample was submitted to Vg=5V for a given waiting time ($t_w$), after which Vg was set back to zero. The filling of the dip which had appeared during $t_w$ at Vg=5V was then measured as the time evolution of G(2.5V)-G(5V). The point at Vg=2.5V is situated between the 0V and 5V dips, and gives a reference G value outside any dip. Comparing curves obtained for different $t_w$ values, one clearly sees an ageing effect: the longer the 5V dip was dug, the slower its refilling is. Finally, it was also observed, like in [6], that an abrupt change of temperature induces a slow logarithmic relaxation of the sample's conductance.

The existence of a symmetrical field effect is not trivial, since for the low field values involved, one would expect a small anti-symmetrical linear effect due to charging the sample. The results on granular Al are strikingly similar to the ones reported on insulating InOx films. Consider first the dip characteristics (Fig. 2). Its amplitude is δG/G~2-3%, like in InOx films of similar R [1]. Its FWHM, when expressed in charge of the gate capacitance ($Q \sim 5 \cdot 10^{11}$ e/cm$^2$) is the same as the value reported for highly doped (non steochiometric) InOx [4]. The absence of a visible anti-symmetrical contribution is consistent with the higher carrier density of the granular Al samples. I observed that the dip amplitude $\delta G$ increases (from $5 \cdot 10^{-9}$ ohm$^{-1}$ to more than $6 \cdot 10^{-8}$ ohm$^{-1}$) when T is increased from 4K to 9K, while $\delta G/G$ decreases (down to 0.5%), again like in InOx in the same temperature range [3]. The relaxation phenomena are also very reminiscent of the InOx behaviour. The two dip experiment (Fig. 3) shows that one can dig a dip to another Vg value, and that the relaxation times involved can be very long (hours or days) like in highly doped InOx. Moreover, the very same ageing effects are observed (Fig. 4), including the simple scaling of the curves with (t/t$_w$) [5] as shown in the inset of Fig. 4. I thus conclude that insulating granular Al-Al$_2$O$_3$ films exhibit the same glassy phenomena as InOx, implying that very similar physics is involved in both systems. This suggests that the phenomena observed are quite general, and can probably equally be observed in many other disordered insulators.

The origin of the above phenomena is not yet clearly identified. In the context of the InOx studies, it was explained on general grounds that when the system is put out of equilibrium by an external perturbation its conductance increases [2], the measure of $\delta G(t)$ then allowing to follow the relaxation to the new equilibrium state. Detailed discussions were given in the papers by Vaknin et al. and a heuristic model was developed which reproduces several experimental features [11]. However the microscopic mechanisms involved are difficult to identify, and the connection of the G(Vg) dip with the Coulomb gap, as suggested in [12], is not yet established. The study of the same phenomena in granular systems may help to answer these questions, as the microstructural parameters which govern the properties can be controlled. The effect of grains charging was invoked by Adkins et al. [7] in the case of Au granular samples, but it was believed to be totally smeared out by the potential disorder created by charge exchanges between the grains and the substrates. It was rather suggested that the appearance and pinning of the G(Vg) dip was due to relaxation phenomena of charges present inside the substrate. However the studies on InOx have shown that the relaxation phenomena correlate with the InOx film's characteristics, which strongly suggests that they are intrinsic to it and not to the insulating substrates.

The granularity of the present sample obviously plays a role and it is tempting to consider the hypothesis that the phenomena are related to charging effects. Charges freshly injected in isolated grains by applying Vg could represent the "bare particles" of the model of [11] responsible for the extra conductance, the relaxation to equilibrium (formation of the "quasiparticles") proceeding by intergrain hops, which are governed by the neighbouring tunnel barriers and capacitances. The equilibrium state of the sample for a given Vg would correspond to the repartition of the charges between the grains which minimizes the energy (and the conductance) of the system. Within this picture one would expect that to increase G out of the dip one has to charge a significant fraction of the grains (by one or several charge quanta, depending on the grains). Then the grain density of the sample and the G(Vg) dip width (expressed as a surface electron density as above) should be comparable. This seems to be the case. It is difficult to determine the grain size in the granular Al channels as some roughness also arises from the underlying Al and Al$_2$O$_3$ thicker layers. But since test granular Al strips were also deposited directly on the sapphire substrates simultaneously with the channels, we could estimate from STM pictures a grain density $n_G \geq 1.6 \cdot 10^{11}$ cm$^{-2}$, a mean grain size around 180 Å and a rather broad distribution spanning from 30 Å to 300 Å. One can expect the grain sizes of the channel deposited on the disordered Al$_2$O$_3$ gate insulator to be similar or smaller. Hence, $\Delta Q$ and $n_G$ are of the same order of magnitude. Clearly this is not a proof and further studies are necessary to test the scaling of the G(Vg) dip width with the grain density. Note that in principle one expects oscillations of G(Vg) stemming from the discrete charging of small grains, but in our case their absence can be attributed to the broad grain size distribution.

Pushing the hypothesis further we can ask whether a similar mechanism could be involved in InOx. Oscillations were observed beside the main dip in G(Vg) curves of highly resistive films [1], and measurements performed on small wires demonstrated single electron charging effects [13]. The occurrence of granular like effects thus seems to be established in this system. Note that both the oscillation period of small wires (involving only one or a few "grains") and the width of the dip and oscillatory structures of the strongly insulating films give similar estimates of the "grain" sizes. The granular scenario in InOx was already considered by the authors, but as they noted its origin is unclear as no obvious granularity appears in the microstructure of the continuous amorphous samples.

In conclusion I have performed conductivity and field effect measurements on insulating granular Al thin films. A symmetrical field effect and very slow conductance relaxations have been observed. I have shown that they are identical to the phenomena already reported in insulating continuous InOx thin films. It shows that these phenomena are not limited to InOx but are probably rather general in disordered insulating systems. It also leads to ask the question of the role of granular effects, where intergrain electron tunnelling and single grain charging are the basic ingredients. Granular films constitute model systems which open a promising route to understand the mechanism of the electron glassy behaviour. In particular, studying samples with controlled tunnel barriers and grain sizes should enable to test the above speculations and distinguish the roles played by the electron localisation and correlations in the phenomena.

I wish to thank Z. Ovadyahu for enthusiastic discussions on the glassy behaviour of InOx, as well as J. Y. Veuillen for discussions and helpful comments. A. Barbara is thanked for the STM examination of the sample, and F. Gay for his advices on the measurement setup.

**Figure caption**

**Fig. 1**: Temperature dependence of the resistance of the granular Al film with R (4K)~280 MΩ. The contacts geometry is such that R ~ R /80.

**Fig. 2**: Gate voltage (Vg) dependence of the conductance G measured at 4K for the sample equilibrated at Vg=0. The Vg scan duration is 5 min.

**Fig. 3**: Two dip experiment performed at 4K. The upper curve is for the sample equilibrated at Vg = 0 V, the lower curves (shifted for clarity) are measured at time intervals of 65' after switching to 1 V. Each scan duration is 5 min. This illustrates how a new dip progressively appears centred on the new Vg value as the initial equilibrium dip vanishes.

**Fig. 4**: Ageing experiment. The curves show the decreasing depth of the G(Vg) dip which was formed by subjecting a sample equilibrated at Vg=0 V to Vg=5 V for a given time $t_w$. From top to bottom, $t_w$'s are: 13 hours, 6 hours, 3 hours and 0.75 hour. The longer the 5 V dip was dug, the slower its vanishing is. The inset shows the simple scaling law: when plotted versus $t/t_w$ all the curves collapse on a single one.

**Figures**

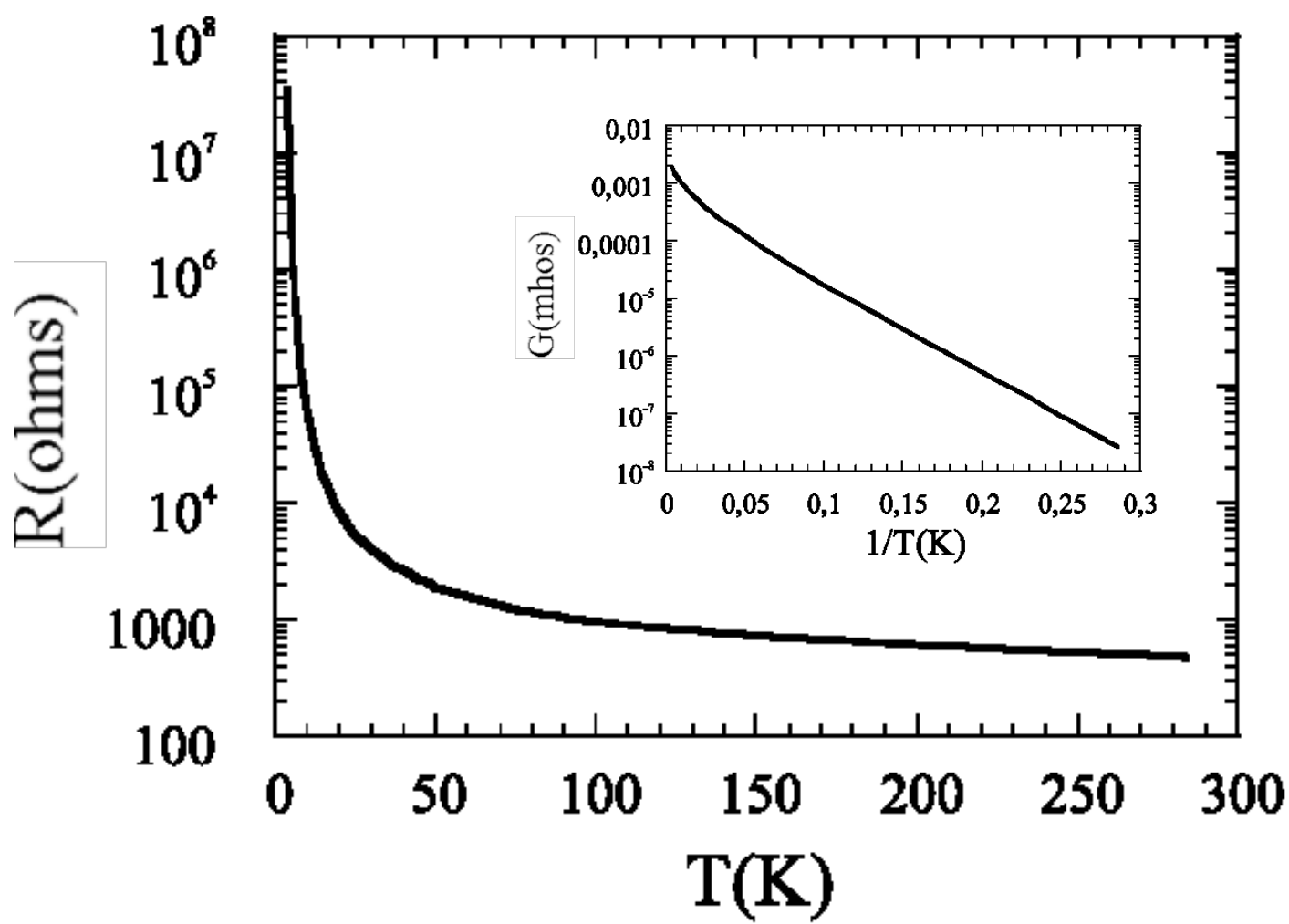

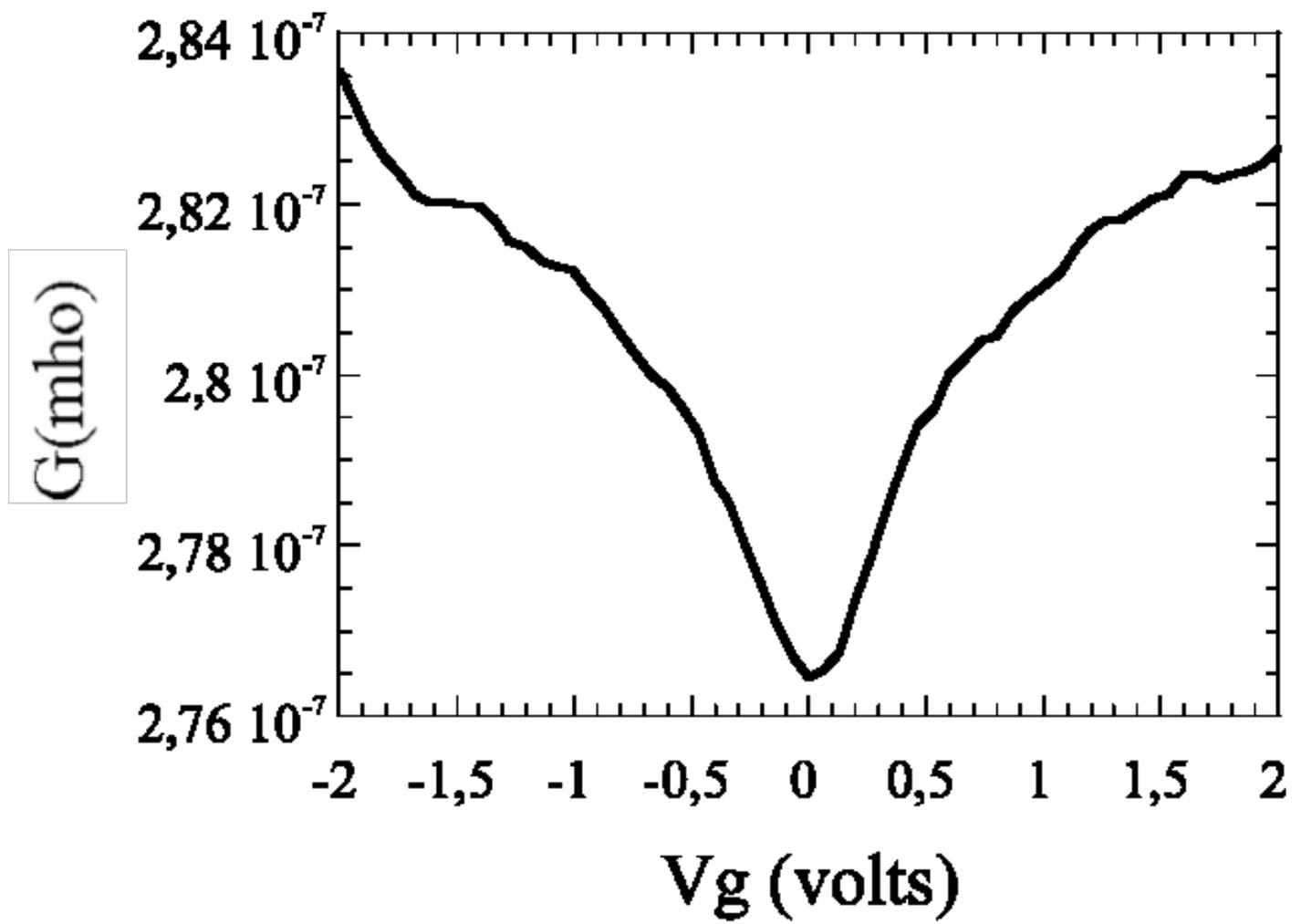

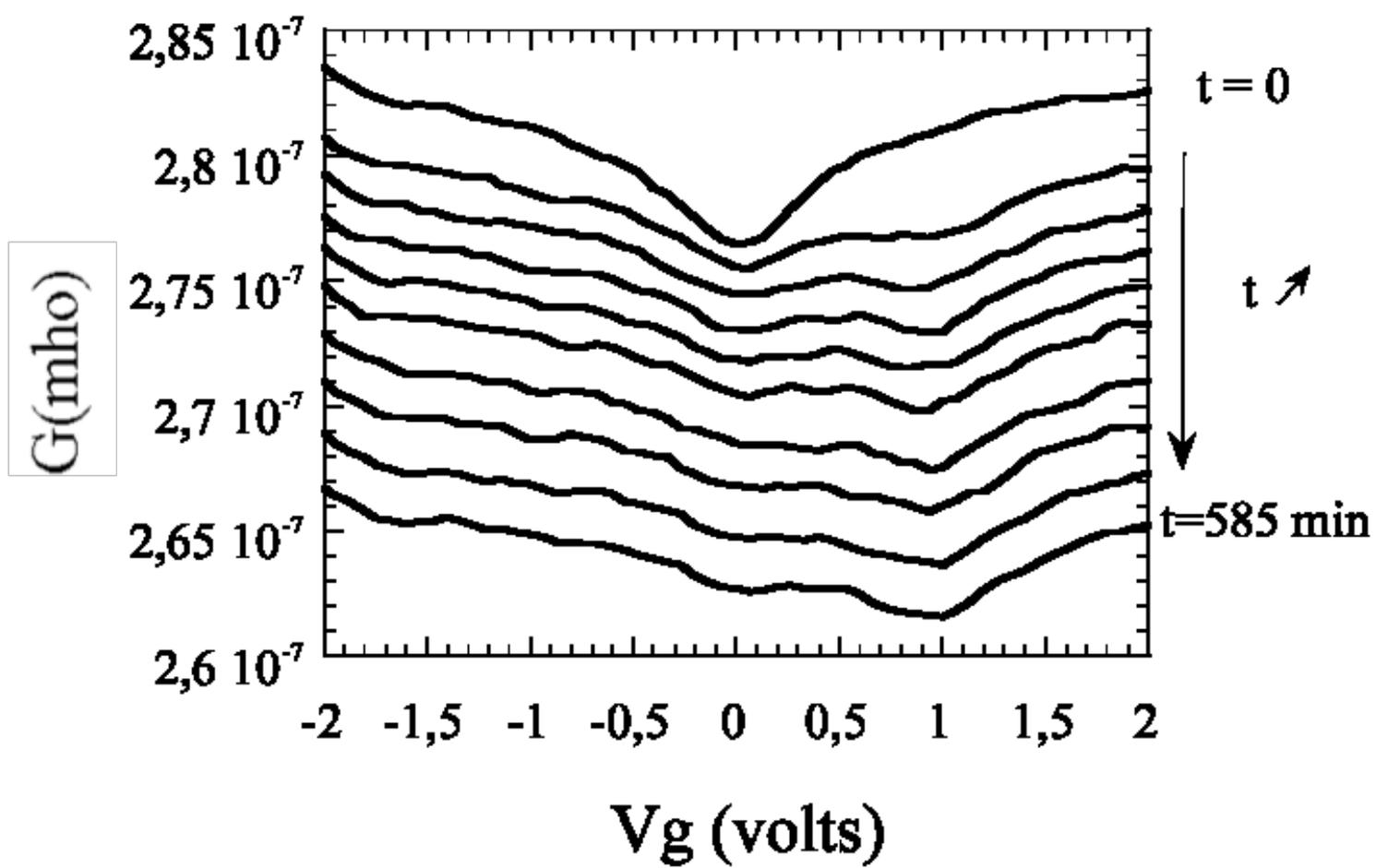

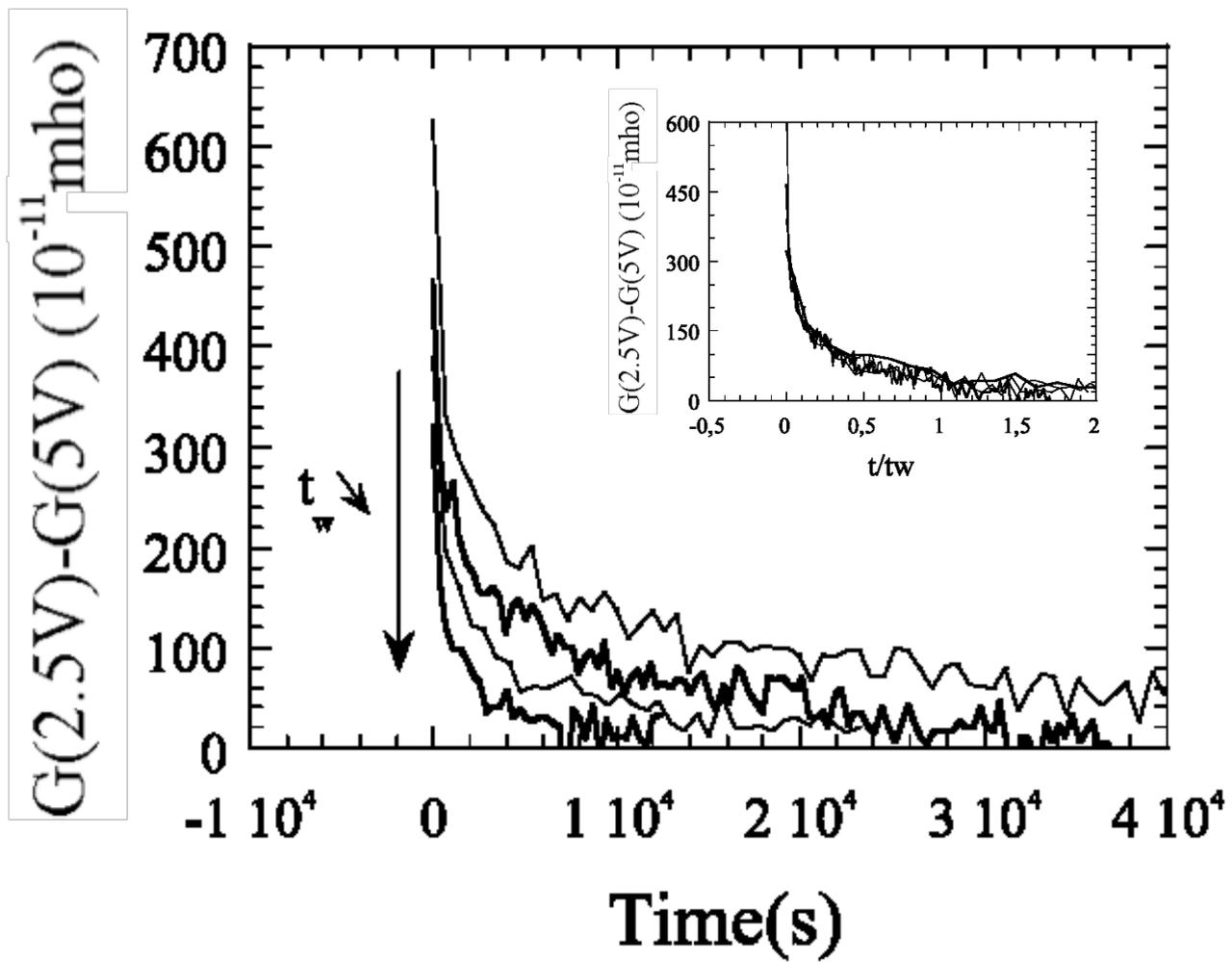